\documentclass[12pt]{iopart}
\usepackage{iopams}
  \usepackage{amssymb}
  \usepackage{graphicx}
  \usepackage{overunderset}

\newfont{\myeu}{eurm10 at 12 pt}
\newfont{\mycal}{eufb10 at 12pt}
\newfont{\bfrak}{eufb10 at 12 pt}
\newcommand{\be}{\begin{equation}}
\newcommand{\ee}{\end{equation}}
\newcommand{\ba}{\begin{eqnarray}}
\newcommand{\ea}{\end{eqnarray}}

\def\bZ{{\bf Z}}
\def\bX{{\bf X}}
\def\bq{{\bf q}}
\def\bI{{\bf I}}
\def\bR{{\bf R}}
\def\bA{{\bf A}}

\def\bC{{\bf C}}

\def\balpha{{\boldsymbol{\alpha}}}
\def\bbeta{{\boldsymbol{\beta}}}
\def\bgamma{{\boldsymbol{\gamma}}}
\def\bGamma{{\boldsymbol{\Gamma}}}
\def\bH{\mbox{\boldmath$\mathsf{H}$}}
\def\bK{{\bf K}}
\def\bmu{{\boldsymbol{\mu}}}
\def\bnu{{\boldsymbol{\nu}}}
\def\bOmega{{\boldsymbol{\Omega}}}
\def\bP{\mbox{\boldmath$\mathsf{P}$}}
\def\bQ{{\bf Q}}
\def\btau{{\boldsymbol{\tau}}}
\def\bV{{\bf V}}
\def\bY{{\bf Y}}
\def\bu{{\bf u}}
\def\vp{{\vphantom{i}}}

\def\halfs{\scriptstyle{\frac 1 2}}

\begin{document}
\title[ ]%
{Parafermions in the $\btau_\mathbf{2}$ model}

\author{Helen Au-Yang$^{1,2}$ and Jacques H H Perk$^{1,2}$}

\address{$^1$ Department of Physics, Oklahoma State University, \\
145 Physical Sciences, Stillwater, OK 74078-3072, USA \\
$^2$ Wuhan Institute of Physics and Mathematics, \\
Chinese Academy of Sciences, Wuhan 430071, China}
\ead{perk@okstate.edu, helenperk@yahoo.com}
\begin{abstract}
It has been shown recently by Baxter that the
$\tau_2(t_q)$ model with open boundary conditions can be solved by the
``parafermionic'' method of Fendley. In Baxter's paper there are several
conjectures, which were formulated based on numerical short-chain
calculations. Here we present the proof of two of them.
\end{abstract}

\section{Introduction}

Parastatistics, more general than Bose or Fermi statistics, has been
advocated first by Green \cite{Green} in 1953, albeit that some
form of generalized statistics was already present implicitly in the
Bethe Ansatz paper \cite{Bethe} of 1931. In 1967, the mathematician
Morris introduced a generalization of the classical Clifford algebra
\cite{Morris1,Morris2,Morris3},\footnote{Yamazaki \cite{Yamazaki},
Popovici and Gh\'eorghe \cite{PG} wrote about this algebra without
giving an explicit representation.} which can be used to describe
`cyclic' parafermions with statistics very different from Green's.
It may also have to be noted that the special case of the Morris
algebra with only two generators is known as the Weyl pair
\cite{Weyl}\footnote{Sylvester introduced this already in 1883
in his paper \cite{Syl} on quaternions, nonions, sedenions, etc.}.
In 1980 Fradkin and Kadanoff \cite{FK} proposed that clock-type
models in two dimensions are an ideal laboratory to study such
parafermions, generated via short-distance expansion of the product
of order and disorder variables, thus generalizing ideas of Kadanoff
and Ceva for the Ising model \cite{KC}.

Many papers have followed since \cite{FK} appeared, including many
papers on the $N$-state chiral Potts model. We should particularly
mention two papers by Fendley \cite{Fen912,Fen1212} on the chiral
Potts quantum spin chain and its specialization for $N=2$, the
quantum Ising chain. In these two papers, the parafermion
operators---introduced in section 3.3 of \cite{Fen912}---are almost
identical in form with the $E$ operators of Morris
\cite{Morris1,Morris2,Morris3}. However, unlike the Ising case,
commuting the Hamiltonian of a chain of length $L$ with a linear
combination of $2L$ of such parafermions does not give rise to
another linear combination of such parafermions \cite{Fen912}.
Fendley \cite{Fen13} considered next the `simple' Hamiltonian
introduced by Baxter \cite{Baxterham} and he constructed $NL$ cyclic
raising operators (called shift operators by him) \cite{Fen13},
allowing him to obtain the complete spectrum of the Hamiltonian.

Baxter constructed his simple Hamiltonian as a special limit
\cite{Bax89} of what is now known as the $\tau_2$ model, which he
obtained specializing the parameters in the integrable chiral Potts model
\cite{BPA}. The existence of this $\tau_2$ model was already implicit
in two papers \cite{Kri81,Kri82} by Krichever, who derived the
genus of the underlying curve as $(N-1)^2$, but only gave explicit
details for case $N=2$, in which he derived the full free-fermion
model as a descendant. Korepanov continued Krichever's program, and
discovered the $\tau_2$ model for $N=3$ in 1986 and for general
$N$ in 1987 \cite{Kor86}, explicitly giving the curve with genus
$(N-1)^2$. Unfortunately, Korepanov's work was  only known to
very few people in Russia until about 1993.

The $\tau_2$ model became really important through the work
of Bazhanov and Stroganov \cite{BS}, who obtained the chiral
Potts model as a descendant of the six-vertex model. They showed
that the $R$-matrix (20) in \cite{BPA}, which is made up of four
chiral Potts model weights, intertwines two $\tau_2$ $R$-matrices.
More precisely they found a sequence of four Yang--Baxter equations,
in which each of the three rapidity lines could either have a
six-vertex model rapidity or two chiral Potts model rapidities.
The name $\tau_2$ came up when the authors of \cite{BBP} decided
to introduce the $\tau_n$ model, whose $R$-matrix intertwines 
a spin-$S$ highest-weight representation and a cyclic representation
while $n=2S+1$; more simply said, one rapidity line belongs to
the spin-$S$ generalization of the six-vertex model and the other
carries two chiral Potts model rapidities. Ever since the two papers
\cite{BS,BBP} most calculations in the integrable chiral Potts
model \cite{BPA,AMPTY} have used the $\tau_2$ model at one point
or another, exploiting the commutation properties of the transfer
matrices of the two models as implied by the Yang--Baxter equations.

Recently Baxter \cite{BaxterPf} generalized Fendley's method to an
inhomogeneous $\tau_2(t)$ model with open boundary conditions. Its
transfer matrix $\btau_2(t)$ is known to be a polynomial in $t$ of
degree $L$,  $\btau_2(t)=\sum_{n=0}^L (-\omega t)^n \bA_n$
and the Hamiltonian is given as $\mathcal{H}=-\bA_1/\bA_0$.
All operators here are explicitly defined in the next section.
In (B4.1) and (B4.2)\footnote{Equations in \cite{BaxterPf} are
denoted here by prefacing B to the equation number.}, Baxter defined
iteratively\footnote{We note that (33) and (34) in \cite{Fen13},
(4) and (1) in \cite{Baxterham}, (8.20) and (8.21) in \cite{Bax89},
and (B1.2) in \cite{BaxterPf} are consistent with one another
identifying $\sigma_m\equiv\bZ_m$ and $\tau_m\equiv\bX_m$.
However, Hamiltonian (B1.5) is spatially reflected compared
to the other papers. Therefore, we must choose (B4.1) and not
$\bGamma_0=\bZ_1$ as would agree with $\psi_1=\sigma_1$ chosen
in subsection 5.1 of \cite{Fen13}.}
\be
\bGamma_0=\bZ_1^{-1},\quad  
\bGamma_{j+1}=(\omega^{-1}-1)^{-1}(\mathcal{H}\bGamma_j-\bGamma_j\mathcal{H}).
\label{gamma0j}\ee
Based on numerical evidence, Baxter found (B4.3), i.e.\ that
\be
s_0 \bGamma_{NL+j}+s_1 \bGamma_{NL-N+j}+\cdots+s_L \bGamma_{j}
=0,\quad \hbox{for $j\ge0$,}
\label{B4.3j}\ee
holds, so that at most $NL$ of the $\bGamma_j$ are linearly
independent, allowing us to truncate the infinite sequence of them.
Furthermore, Baxter showed that there exists a linear transformation
to transform the $NL$ operators $\bGamma_j$ to cyclic raising operators
${\widehat\bGamma}_j$. He showed that these operators are to satisfy
(B5.2) and (B5.4), as conjectured based on numerical evidence for
spin chains of length up to 6.

Finally, to obtain the spectrum of $\tau_2(t)$, he defined in (B4.7)
\be
\bmu_j\equiv\bGamma_j\btau_2(t)-\btau_2(t)\bGamma_j,\quad
\bnu_j\equiv\omega\bGamma_j\btau_2(t)-\btau_2(t)\bGamma_j,
\label{dfmunu}\ee
and observed numerically that $t\bnu_j=\bmu_{j-1}$ in (B4.8). One
expects some relation of this type replacing the $\bGamma_j$ in the
$\nu_j$ by $\bGamma_{j-1}$ upon using (\ref{gamma0j}). The definition
of $\nu_j$ has an $\omega$-commutator and this must be seen as a
consequence of the denominator $(\omega^{-1}-1)$ in (\ref{gamma0j}).
That this has to be so can be most easily seen expanding the special
case $t\bnu_1=\bmu_0$ to linear order in $t$ using
$\btau_2(t)=\mathbf{1}+\omega t\mathcal{H}+\mathrm{O}(t^2)$.

In this paper we shall present proofs of conjectures (B4.3)
and (B4.8). We shall also simplify (B5.4) and present explicit forms
for the operators. We are using the method of commuting transfer
matrices within the Yang--Baxter approach, which is very different
from the method of Fendley \cite{Fen13}, who uses the method of
iteratively constructing commuting local Hamiltonians. Nevertheless,
his generating function $T(t)$ in his equations (48) and (50) is
simply related to our $\btau_2(t)$ in the special limit (B3.25),
as we shall show in Appendix B.
\section{Transfer matrix and Hamiltonian}

The transfer matrix of the generalized $\tau_2$ model \cite{BaxterFun}
can be written as a product of interaction-round-a-face weights, as
was done in \cite{BaxterPf} based on (14) and Figure 4 of
\cite{BaxterFun}. Alternatively, it can also be written as a product
of vertex-model $\mathcal{L}$-matrices as indicated in Figure 5 of
\cite{BaxterFun}. In fact, equation (20) in \cite{BaxterFun} gives
the $\mathcal{L}$-matrix acting on vector ${\bf g}_J$, so that we can
express the $\tau_2$ transfer matrix as the $2\times2$ trace
\be
\btau_2(t)= {\rm trace}\left(\prod_{j=0}^L\mathcal{L}_j\right).
\label{tau2}\ee
From Appendix A we obtain\footnote{Compare (\ref{Lsj}) with the
action on vector $\mathbf{g}_J$ in (20) of \cite{BaxterFun},
identifying $m_{j-1}=m$, $m_j=m'$, $\sigma_j=a$, $\sigma'_j=d$.
The difference is a factor $(-\omega t_q)^{m-m'}$ corresponding
to a simple gauge transformation.}
\ba\fl
&&\mathcal{L}_j(m_{j-1},m_j;\sigma_j,\sigma'_j)
=\mathcal{L}_j(m_{j-1},m_j)_{\sigma_j,\sigma'_j}\nonumber\\
\fl &&=\omega^{m_j\sigma'_j-m_{j-1}\sigma_j}
(-\omega t_q)^{\sigma_j-\sigma'_j-m_{j-1}}
F_{2j-2}(\sigma_j-\sigma'_j|m_{j-1})F_{2j-1}(\sigma_j-\sigma'_j|m_{j}),
\label{Lsj}\ea
using (\ref{Vpqq'}), (\ref{Vpq'q}) and (\ref{Lpp'q}).
Also, we must identify
$F_{2j-2}(n|m)=F_{p_{2j-2},q}(n,m)$ and
$F_{2j-1}(n|m)=F_{p_{2j-1},q}(n,m)$ when comparing
with (B2.2).
Rewriting the $\mathcal{L}_j$ as 2-by-2 matrices with $N$-by-$N$
matrix elements, we thus find
\ba
\mathcal{L}_j(0,0)=
b_{2j-2}b_{2j-1}-\omega t_q d_{2j-2}d_{2j-1}\bX_j,\nonumber\\
\mathcal{L}_j(0,1)=
(-\omega t_q)\bZ_j(b_{2j-2}c_{2j-1}-d_{2j-2}a_{2j-1}\bX_j),\nonumber\\
\mathcal{L}_j(1,0)=
\bZ_j^{-1}(c_{2j-2}b_{2j-1}-\omega a_{2j-2}d_{2j-1}\bX_j),\nonumber\\
\mathcal{L}_j(1,1)=
\omega a_{2j-2}a_{2j-1}\bX_j-\omega t_q c_{2j-2}c_{2j-1},
\label{Lj}\ea
where
\ba
[\bZ_j]_{\sigma,\sigma'}=\omega^{\sigma_j}
\prod_{k=0}^L\delta(\sigma^\vp_k, \sigma'_k),
\quad
[\bX_j]_{\sigma,\sigma'}=\delta(\sigma^\vp_j,\sigma'_j+1)
\prod_{k\ne j}\delta(\sigma^\vp_k, \sigma'_k),\nonumber\\
 \bZ_j\bX_j=\omega\bX_j\bZ_j.
\label{morris}\ea
Particularly, for $c_{2L}\equiv c_{-2}=c_{-1}=0$,
$a_{-1}=d_{-1}=0$ and $b_{-1}=b_{-2}=1$, in agreement
with (B3.1), (B3.4) and (B3.6),  we find 
\be
\mathcal{L}_0=
\left[\begin{array}{cc}
{\bf 1} &0\\
0 &0\end{array}\right].
\label{L0}
\ee
Let 
\be
\prod_{j=1}^L\mathcal{L}_j=\left[\begin{array}{cc}
{\bf A} (t)&{\bf B} (t)\\
{\bf C} (t) &{\bf D} (t)\end{array}\right],
\label{ABCD}\ee
then from (\ref{tau2}), (\ref{L0}), and (\ref{ABCD}), we find
\be
\btau_2(t)={\bf A}(t)=\sum_{\ell=0}^L {\bf A}_\ell(-\omega t)^\ell,
\label{A}\ee
where the ${\bf A}_\ell$ are operators commuting with one another.
Indeed $[\tau_2(t),\tau_2(t')]=0$  as follows from Yang--Baxter
equation (\ref{YBE3}), which is valid for all inhomogeneous choices
of the rapidities $p_j=\{a_j,b_j,c_j,d_j\}$ \cite{BaxterFun}.
Next we rewrite (\ref{Lj}) as
\be
\mathcal{L}_j=\mathcal{L}^+_j-\omega t\mathcal{L}^-_j,
\label{Ljsum}\ee
where the $\mathcal{L}^+_j$ and $\mathcal{L}^-_j$ are both triangular,
\be
\mathcal{L}^+_j=\left[\begin{array}{cc}
\balpha_j^+&0\\
\bbeta_j^+&\bgamma^+_j\end{array}\right],
\quad
\mathcal{L}^-_j=\left[\begin{array}{cc}
\balpha_j^-&\bbeta_j^-\\
0&\bgamma^-_j\end{array}\right],
\label{Ljpm}
\ee
and respectively given by the constant terms or the linear terms
in (\ref{Lj}). Consequently, we find\footnote{We do not set
$b_j\equiv1$ as done in \cite{BaxterPf}, so that we can treat the
superintegrable case later.}
\be
{\bf A}_0=\prod_{j=1}^L\balpha^+_j=
\Bigg[\prod_{j=0}^{2L-1}b_j\Bigg]{\bf1},
\quad 
 {\bf A}_L=\prod_{j=1}^L\balpha^-_j
 =\Bigg[\prod_{j=0}^{2L-1}d_j\Bigg]\bX_1\cdots\bX_L,
\label{A0L}\ee
with ${\bf A}_0=A_0{\bf1}$ and $\balpha_j^+=\alpha_j^+{\bf1}$
proportional to the unit operator, and the Hamiltonian 
\be
\mathcal{H}=-\frac{{\bf A}_1}{A_0}=
-\sum_{j=1}^L\Bigg[\frac{\balpha^-_j}{\alpha^+_j}
+\frac{\bbeta^-_j}{\alpha^+_j}
\sum_{m=j+1}^L\Bigg(\prod_{\ell=j+1}^{m-1}\frac{\bgamma^+_\ell}
{\alpha^+_\ell}\Bigg)
\frac{\bbeta^+_m}{\alpha^+_m}\Bigg],
\label{H}
\ee
can be easily shown to be identical to (B3.22). It should be noted
that this Hamiltonian is not the one of the integrable chiral
Potts chain \cite{AMPTY} as studied by Fendley \cite{Fen912,Fen1212},
but it reduces in the special limit (B3.25) to the one he studied in
\cite{Fen13}, see appendix B.

Since, from (B4.1),
\be
\bGamma_0=\bZ_1^{-1},
\label{Gamma0}\ee
we may split the product in (\ref{ABCD}) into two parts
\be
\prod_{j=1}^L\mathcal{L}_j=\mathcal{L}_1\prod_{j=2}^L\mathcal{L}_j,
\ee
and rewrite the second part as
\be
\prod_{j=2}^L\mathcal{L}_j=
\left[\begin{array}{cc}{\bf A}^{2,L} (t)&{\bf B}^{2,L} (t)\\
{\bf C}^{2,L} (t) &{\bf D}^{2,L} (t)\end{array}\right],
\ee
which makes explicit that it is a $2\times2$ matrix with
operator entries. It follows that
\be
\btau_2(t)={\bf A}(t)=(\balpha^+_1-\omega t\balpha^-_1)
{\bf A}^{2,L} (t)-\omega t\bbeta^-_1{\bf C}^{2,L} (t),
\label{relationA}\ee
where
\be
\balpha^+_1=b_0b_1{\bf 1},\quad \balpha^-_1=d_0d_1\bX_1,\quad
\bbeta^-_1=\bZ_1(b_0c_1-d_0a_1\bX_1).
\label{alpha}\ee
Expanding
\be
{\bf A}^{2,L} (t)=
\sum_{\ell=0}^{L-1}{\hat{\bf A}}_\ell (-\omega t)^\ell,\quad
{\bf C}^{2,L} (t)=
\sum_{\ell=0}^{L-1}{\hat{\bf C}}_\ell (-\omega t)^\ell,
\label{hatAC}\ee
and substituting this and (\ref{A}) into (\ref{relationA}),
we can relate the coefficients as
\be
{\bf A}_\ell=
\balpha^+_1{\hat{\bf A}}_\ell+\balpha^-_1{\hat{\bf A}}_{\ell-1}
+\bbeta^-_1{\hat{\bf C}}_{\ell-1}.
\label{Al}\ee
Particularly, the Hamiltonian (\ref{H}) can be rewritten as
\be 
-{\bf A}_0\mathcal{H}={\bf A}_1=
\balpha^+_1{\hat{\bf A}}_1+\balpha^-_1{\hat{\bf A}}_{0}
+\bbeta^-_1{\hat{\bf C}}_{0}.
\label{A1}\ee
Obviously, as ${\hat{\bf A}}_\ell$ and ${\hat{\bf C}}_\ell$ are
operators acting on sites from 2 to $L$, they commute with $\bGamma_0$,
$\balpha_1^\pm$ and $\bbeta_1^-$. Using this, the iterative definition
(B4.2), i.e.
\be 
\bGamma_j{\bf A}_1-{\bf A}_1\bGamma_j=(\omega^{-1}-1)
{\bf A}_0\bGamma_{j+1},
\label{gamma}\ee
the third equation (\ref{morris}) rewritten as
$\bX_1\bGamma_0=\omega\bGamma_0\bX_1$,
and (\ref{A1}), we find
\be 
{\bf A}_0\bGamma_{1}=
\omega(\bGamma_0\balpha_1^-{\hat{\bf A}}_0-d_0a_1\bX_1{\hat{\bf C}}_0).
\label{gamma1}\ee
We shall first prove (B4.8), which is the easiest.
\section{Proof of (B4.8)}

From the definitions of $\bmu_j$ and $\bnu_j$ in (\ref{dfmunu}),
cf. (B4.7), and the definition of the Gamma matrices in (\ref{gamma0j}),
we find
\be
\mathcal{H}\bmu_j-\bmu_j\mathcal{H}=\bmu_{j+1}(\omega^{-1}-1),\quad
\mathcal{H}\bnu_j-\bnu_j\mathcal{H}=\bnu_{j+1}(\omega^{-1}-1).
\label{hmn}\ee
Thus if we can prove $t\bnu_1=\bmu_0$, then by repeated application
of (\ref{hmn}) on both sides, we can obtain  $t\bnu_{j+1}=\bmu_j$.
Using the expansion in (\ref{A}) and the definitions (\ref{dfmunu}),
we find
\ba
t\bnu_1=\sum_{\ell=1}^{L}(-\omega t)^\ell
(\omega^{-1}{\bf A}_{\ell-1}\bGamma_1-\bGamma_1{\bf A}_{\ell-1}),
\nonumber\\
\bmu_0=\sum_{\ell=1}^{L}(-\omega t)^\ell
(\bGamma_0{\bf A}_{\ell}-{\bf A}_{\ell}\bGamma_0).
\ea
If we can prove that the coefficients of $t^\ell$ are identical,
then the identity is proven. It is easily seen from (\ref{gamma})
that this equality holds for $\ell=1$. From (\ref{Al}) and
(\ref{alpha}), we find
\be
\bGamma_0{\bf A}_{\ell}-{\bf A}_{\ell}\bGamma_0=
(1-\omega)(\bGamma_0\balpha^-_1{\hat{\bf A}}_{\ell-1}
-d_0a_1\bX_1{\hat{\bf C}}_{\ell-1}),
\label{4.8a}\ee
and from (\ref{Al}) and (\ref{gamma1}), we obtain
\ba\fl
\omega^{-1}{\bf A}_{\ell-1}\bGamma_1-\bGamma_1{\bf A}_{\ell-1}
\nonumber\\
\fl=A_{0}^{-1}\Big[(1-\omega)
\bGamma_0(\balpha^+_1{\hat{\bf A}}_0)\balpha^-_1{\hat{\bf A}}_{\ell-1}
+(\bbeta^-_1\bGamma_0\balpha^-_1-\omega\bGamma_0\balpha^-_1\bbeta^-_1)
{\hat{\bf A}}_0{\hat{\bf C}}_{\ell-2}\nonumber\\
\fl+\balpha^+_1d_0a_1\bX_1(\omega{\hat{\bf C}}_0{\hat{\bf A}}_{\ell-1}
-{\hat{\bf A}}_{\ell-1}{\hat{\bf C}}_0)
+\balpha^-_1d_0a_1\bX_1(\omega{\hat{\bf C}}_0{\hat{\bf A}}_{\ell-2}-
{\hat{\bf A}}_{\ell-2}{\hat{\bf C}}_0)\Big],
\label{4.8b}\ea
where the relations
\be
\bbeta^-_1\bX_1=\omega\bX_1\bbeta^-_1,\quad
\balpha^-_1\bGamma_0=\omega\bGamma_0\balpha^-_1,
\label{commu}\ee
and (\ref{comCA}) have also been used. 

From the Yang--Baxter equations\footnote{Details will be discussed
in the Appendix, as there are rather subtle differences depending
on the various conventions.} we obtain the relation
\be\fl
\omega^{-1}(1-x/y){\bf A}(y){\bf C}(x)+
(1-\omega^{-1}){\bf C}(y){\bf A}(x)=
(1-\omega^{-1}x/y){\bf C}(x){\bf A}(y).
\label{YBEAC}\ee
By equating the coefficients, we find 
\be 
{\hat{\bf A}}_\ell {\hat{\bf C}}_0-
\omega{\hat{\bf C}}_0{\hat{\bf A}}_\ell=
(1-\omega){\hat{\bf C}}_\ell{\hat{\bf A}}_0=
(1-\omega){\hat{\bf A}}_0{\hat{\bf C}}_\ell,
\label{YB1}\ee
and
\ba 
{\hat{\bf A}}_1 {\hat{\bf C}}_\ell-{\hat{\bf C}}_\ell{\hat{\bf A}}_1
&=&(1-\omega)({\hat\bC}_{\ell+1}{\hat\bA}_0-
{\hat\bC}_0{\hat\bA}_{\ell+1})
\label{YB2}\\
&=&{\hat\bA}_{\ell+1}{\hat\bC}_0-{\hat\bC}_0{\hat\bA}_{\ell+1}
\nonumber\\
&=&(1-\omega^{-1})({\hat\bA}_{\ell+1}{\hat{\bf C}}_0-
{\hat\bA}_0{\hat{\bf C}}_{\ell+1}).
\label{YB3}\ea
To go from (\ref{YB2}) to (\ref{YB3}) via the indicated
intermediate step requires two applications of (\ref{YB1}).
From (\ref{A0L}) and (\ref{alpha}),  we have
\ba
(\balpha^+_1{\hat{\bf A}}_0)={\bf A}_0,\quad
\bbeta^-_1\bGamma_0\balpha^-_1-\omega\bGamma_0\balpha^-_1\bbeta^-_1=
(1-\omega)d_0a_1\balpha^-_1\bX_1.
\label{beta}\ea
There are four terms within the square brackets of (\ref{4.8b}).
Using (\ref{beta}) for the first and second terms and (\ref{YB1})
for the third and fourth terms, one can show that
the right-hand sides of (\ref{4.8a}) and (\ref{4.8b})
are equal, so that
\be
\omega^{-1}{\bf A}_{\ell-1}\bGamma_1-\bGamma_1{\bf A}_{\ell-1}=
\bGamma_0{\bf A}_{\ell}-{\bf A}_{\ell}\bGamma_0
\ee
for all $\ell$. Thus we have proven the identity (B4.8)
in \cite{BaxterPf}.
\section{Proof of (B4.3)}

\subsection[4.1]{Explicit form of $\bGamma_j$}
In (\ref{gamma1}),  $\bGamma_1$ is explicitly given. We shall prove
by induction that for $\ell\ge1$
\ba
\bGamma_\ell=\omega^\ell\sum_{m=0}^{\ell-1}(-1)^m\bR_m \bq_{\ell-1-m}
=\omega\sum_{m=0}^{\ell-1}(-1)^m \bq_{\ell-1-m}\bR_m,
\label{gammaj}\ea
where
\ba
\bR_m\equiv\bGamma_0\balpha^-_1{\hat{\bf A}}_m
-d_0a_1\bX_1{\hat{\bf C}}_m,
\label{Rm}\ea
in which the hatted operators do not commute with the $\bA_m$,
but commute with $\balpha_1^{\pm}$, $\bbeta_1^{-}$ and $\bGamma_0$,
while the $\bq_{\ell}$ are operators which can be obtained
iteratively by the relations
\be
\bq_0=\frac{\bf 1}{A_{0}}, \quad\bq_\ell=
\sum_{n=1}^\ell (-1)^{n+1}\frac{ {\bf A}_{n}}{A_{0}}\bq_{\ell-n}.
\label{q}\ee
Since the $\bq_\ell$ are expressed in terms of the $\bA_n$, they
commute with all $\bA_n$. The second equality in (\ref{gammaj}) is
needed only for the next section.

Comparing (\ref{gammaj}) with (\ref{gamma1}), we find it gives
the right result for $\bGamma_1$. Now we assume (\ref{gamma1})
holds for $\bGamma_\ell$, and prove it is also correct for
$\bGamma_{\ell+1}$. Using (\ref{gamma}) and (\ref{gammaj}), we find
\ba\fl
(1-\omega^{-1}) {\bf A}_{0}\bGamma_{\ell+1}=
{\bf A}_{1}\bGamma_\ell-\bGamma_\ell{\bf A}_{1}
&=&\omega^\ell\sum_{m=0}^{\ell-1}(-1)^m
(\bA _1\bR_m-\bR_m\bA_1) \bq_{\ell-1-m},
\label{gammajj}\\
&=&\omega\sum_{m=0}^{\ell-1}(-1)^m\bq_{\ell-1-m}
(\bA _1\bR_m-\bR_m\bA_1).
\label{gammajjj}\ea
Using (\ref{Rm}) we split the commutator $\bA _1\bR_m-\bR_m\bA_1$
into two parts ${\bf I}_1-{\bf I}_2$, with
\be\fl
{\bf I}_1={\bf A}_{1}\bGamma_0\balpha^-_1{\hat{\bf A}}_m-
\bGamma_0\balpha^-_1{\hat{\bf A}}_m{\bf A}_{1},\quad
{\bf I}_2=d_0a_1({\bf A}_{1}\bX_1{\hat{\bf C}}_m-
\bX_1{\hat{\bf C}}_m{\bf A}_{1}).
\label{bI}\ee
After substituting (\ref{A1}) into (\ref{bI}),
we use the commutation relations (\ref{commu}) and (\ref{comCA}) and
the fact that the hatted operators commute with all operators on
site 1 and find
\ba\fl
{\bf I}_{1}&=&
(\omega-1)\bGamma_0(\balpha^-_1)^2{\hat{\bf A}}_0{\hat{\bf A}}_m+
\balpha^-_1\big[\bbeta_1^-\bGamma_0{\hat{\bf C}}_0{\hat{\bf A}}_m-
\omega^{-1}\bGamma_0\bbeta_1^-{\hat{\bf A}}_m{\hat{\bf C}}_0\big]
\label{I1a}\\ \fl
&=& (\omega-1)\bGamma_0(\balpha^-_1)^2{\hat{\bf A}}_0{\hat{\bf A}}_m
\nonumber\\ \fl
&&+\balpha^-_1\bigg[(\bbeta_1^-\bGamma_0-\bGamma_0\bbeta_1^-)
{\hat{\bf C}}_0{\hat{\bf A}}_m+\bGamma_0\bbeta_1^-
({\hat{\bf C}}_0{\hat{\bf A}}_m-
\omega^{-1}{\hat{\bf A}}_m{\hat{\bf C}}_0)\bigg].
\label{I1a2}\ea
Next we use  (\ref{YB1})  and combining (\ref{alpha}) and (\ref{Gamma0})
we may write
\be
\bbeta_1^-\bGamma_0-\bGamma_0\bbeta_1^-=-(\omega-1)d_0a_1\bX_1,
\label{comb}\ee
so that
\ba\fl
{\bf I}_{1}&=&
(1-\omega^{-1})
\bigg[\omega\bGamma_0(\balpha^-_1)^2{\hat{\bf A}}_0{\hat{\bf A}}_m
+\balpha^-_1(\bGamma_0\bbeta^-_1{\hat{\bf C}}_m{\hat{\bf A}}_0-
\omega d_0a_1\bX_1{\hat{\bf C}}_0{\hat{\bf A}}_m)\bigg]\nonumber\\
\fl &=&(1-\omega^{-1})\bigg[\omega\bGamma_0\balpha^-_1
(\balpha^-_1{\hat{\bf A}}_m+\bbeta^-_1{\hat{\bf C}}_m){\hat{\bf A}}_0-
\omega \balpha^-_1d_0a_1\bX_1{\hat{\bf C}}_0{\hat{\bf A}}_m\bigg],
\ea
from which, using (\ref{Al}) and (\ref{beta}), we obtain
\ba
{\bf I}_{1}
=(1-\omega^{-1})\omega\bigg[\bGamma_0\balpha^-_1
({\hat{\bf A}}_0{\bf A}_{m+1}-{\hat{\bf A}}_{m+1}{\bf A}_0)
-\balpha^-_1d_0a_1\bX_1{\hat{\bf C}}_0{\hat{\bf A}}_m\bigg].
\label{I1}\ea
Similarly, we find, using (\ref{A1}), (\ref{commu}) and (\ref{comCA}),
\ba
{\bf I}_{2}=d_0a_1\bigg[\balpha^+_1\bX_1({\hat{\bf A}}_1{\hat{\bf C}}_m
-{\hat{\bf C}}_m{\hat{\bf A}}_1)+
(\omega-1)\bX_1\bbeta^-_1{\hat{\bf C}}_0{\hat{\bf C}}_m\bigg],
\label{I2a}\ea
which, upon using first (\ref{YB2}) and in the next step (\ref{beta})
and (\ref{Al}), becomes
\ba
{\bf I}_{2}&=&-(1-\omega^{-1})\omega d_0a_1\bX_1
\bigg[\balpha^+_1{\hat{\bf C}}_{m+1}{\hat{\bf A}}_0
-{\hat{\bf C}}_0(\balpha^+_1{\hat{\bf A}}_{m+1}
+\bbeta^-_1{\hat{\bf C}}_m)\bigg]\nonumber\\
&=&-(1-\omega^{-1})\omega d_0a_1\bX_1
({\hat{\bf C}}_{m+1}{\bf A}_0-{\hat{\bf C}}_0{\bf A}_{m+1}+
\balpha^-_1{\hat{\bf C}}_0{\hat{\bf A}}_m).
\label{I2}\ea
Combining (\ref{I1}) and (\ref{I2}), we find that the last terms in
the two equations cancel out. The definition of $\bR_m$ in (\ref{Rm})
is then used to write
\ba
[\bA_1,\bR_m]=\bI_1-\bI_2=
(1-\omega^{-1})\omega (\bR_0\bA_{m+1}-\bR_{m+1}A_0).
\label{comR1}\ea
Substituting (\ref{comR1}) into (\ref{gammajj}), we find
\ba
\bGamma_{\ell+1}=\omega^{\ell+1}
\bigg[ \bR_0\sum_{m=0}^{\ell-1}(-1)^m
\frac{{\bf A}_{m+1}}{A_0}\bq_{\ell-1-m}-
\sum_{m=0}^{\ell-1}(-1)^m\bR_{m+1}\bq_{\ell-1-m}\bigg].
\ea
Noticing from (\ref{q}) that the coefficient of $\bR_0$ is $\bq_\ell$
and replacing $m$ by $m-1$ in the second sum, we find $\bGamma_{\ell+1}$
is also of the form (\ref{gammaj}), thus completing the proof of the
first equality in (\ref{gammaj}).

Alternatively, we may rewrite (\ref{I1a}) as 
\ba\fl
{\bf I}_{1}&=&
(\omega-1)\bGamma_0(\balpha^-_1)^2{\hat{\bf A}}_0{\hat{\bf A}}_m
\nonumber\\ \fl 
&&+\balpha^-_1\big[\bbeta_1^-\bGamma_0({\hat{\bf C}}_0{\hat{\bf A}}_m-
\omega^{-1}{\hat{\bf A}}_m{\hat{\bf C}}_0)+
\omega^{-1}(\bbeta_1^-\bGamma_0-\bGamma_0\bbeta_1^-)
{\hat{\bf A}}_m{\hat{\bf C}}_0\big].
 \ea
After again using (\ref{comb}) and (\ref{YB1}) and performing a few
commutations with the help of  (\ref{commu}) and (\ref{comCA}),
we can apply (\ref{Al}) to arrive at the alternative form
\ba\fl
{\bf I}_{1}
 =(1-\omega^{-1})
\bigg[{\bf A}_{m+1}\bGamma_0\balpha^-_1{\hat{\bf A}}_0
 -{\bf A}_0\bGamma_0\balpha^-_1{\hat{\bf A}}_{m+1}
 -\balpha^-_1d_0a_1\bX_1{\hat{\bf A}}_m{\hat{\bf C}}_0\bigg].
\label{I1b}\ea
Next, similar to what we did in deriving (\ref{I2}), we now use
commutation relation (\ref{YB3}) followed by (\ref{commu}) and
(\ref{Al}) to rewrite (\ref{I2a}) as
\ba\fl
{\bf I}_{2}
 =(1-\omega^{-1})\bigg[{\bf A}_{m+1}d_0a_1\bX_1{\hat{\bf C}}_0
 -{\bf A}_0d_0a_1\bX_1{\hat{\bf C}}_{m+1}
 -\balpha^-_1d_0a_1\bX_1{\hat{\bf A}}_m{\hat{\bf C}}_0\bigg],
\label{I2b}\ea
so that 
\ba
[\bA_1,\bR_m]=\bI_1-\bI_2=(1-\omega^{-1}) (\bA_{m+1}\bR_0-A_0\bR_{m+1}).
\label{comR2}\ea
Consequently, we find (\ref{gammajjj}) becomes
\ba
\bGamma_{\ell+1}=
\omega\Bigg[\bigg(\sum_{m=1}^{\ell}(-1)^{m+1}\bq_{\ell-m}
\frac{{\bf A}_{m}}{A_0}\bigg)\bR_0
+\sum_{m=1}^{\ell}(-1)^m\bq_{\ell-m}\bR_{m}\Bigg].
\ea
Again we use (\ref{q}) to show that the second equality in
(\ref{gammaj}) also holds for $\ell+1$.

\subsection{Proof of (B4.3)}

We first rewrite (\ref{q}) as
\be
\sum_{n=0}^\ell (-1)^{n} {\bf A}_{n}\bq_{\ell-n}
=\delta_{\ell,0}\mathbf{1}.
\label{q0}\ee
Because ${\bf A}_n=0$ for $n> L$ and $\bq_\ell=0$ for
$-N<\ell<0$,\footnote{Compare eqs.~(50) of \cite{APsu2}
and (71) of \cite{APsu4} and nearby text.}
the upper limit of the summation can be replaced by $L$ or larger.
It is easily seen from (\ref{A}) that
\ba
\prod_{n=0}^{N-1}\btau_2(\omega^nt)
=\sum_{\ell=0}^L s_\ell t^{N\ell}\mathbf{1}
=\prod_{n=0}^{N-1}\Bigg[\sum_{\ell_n=0}^L
\bA_{\ell_n}(-\omega^{n+1}t)^{\ell_n}\Bigg]\nonumber\\
=\sum_{m=0}^{NL}(-t)^m\underset{\ell_1+\cdots+\ell_N=m}{\sum\cdots\sum}
\bA_{\ell_1}\bA_{\ell_2}\cdots\bA_{\ell_N}
\omega^{\ell_1+2\ell_2+\cdots+N\ell_N}.
\label{prodsum}\ea
As it is obvious that
\be
\underset{\ell_1+\cdots+\ell_N=m}{\sum\cdots\sum}
\bA_{\ell_1}\bA_{\ell_2}\cdots\bA_{\ell_N}
\omega^{\ell_1+2\ell_2+\cdots+N\ell_N}=0 \quad
\mbox{for $m\ne jN$},\label{vanish}\ee
we recover (B3.14) with
\be
s_j\mathbf{1}=
(-1)^{jN}\underset{\ell_1+\cdots+\ell_N=jN}{\sum\cdots\sum}
\bA_{\ell_1}\bA_{\ell_2}\cdots\bA_{\ell_N}
\omega^{\ell_1+2\ell_2+\cdots+N\ell_N}.
\label{sj}\ee
Now consider the sum
\be
\bK=\sum_{\ell=0}^L s_\ell \bGamma_{NL-\ell N}.
\label{K}\ee
Substituting (\ref{sj}) into it, and using (\ref{vanish}),
we rewrite it as
\ba\fl
\bK=\sum_{m=0}^{NL}\bGamma_{NL-m}(-1)^m
\underset{\ell_1+\cdots+\ell_N=m}{\sum\cdots\sum}
\bA_{\ell_1}\bA_{\ell_2}\cdots\bA_{\ell_N}
\omega^{\ell_1+2\ell_2+\cdots+N\ell_N}\nonumber\\
\fl
=\sum_{\ell_1=0}^L\cdots\sum_{\ell_N=0}^L
\bGamma_{NL-\ell_1-\cdots-\ell_N}(-1)^{\ell_1+\cdots+\ell_N}
\bA_{\ell_1}\bA_{\ell_2}\cdots\bA_{\ell_N}
\omega^{\ell_1+2\ell_2+\cdots+N\ell_N}.
\ea
Since (\ref{gammaj}) is not valid for $\bGamma_0$, it cannot be
used when $\ell_1=\cdots=\ell_n=L$ in the above $N$-fold sum. Setting
this term, which is easily simplified, apart and denoting the remaining
$(L+1)^N-1$ terms by putting primes on the sums, we split $\bK$ into
two parts. Next we substitute (\ref{gammaj}) into the remaining sum
part, after changing the upper limit of the summation of (\ref{gammaj})
to $L-1$. Because ${\hat{\bf A}}_m={\hat{\bf C}}_m=0$ for $m\ge L$
and $\bq_\ell=0$ for $-L<\ell<0$, the two choices of the upper limits
for $m$ are equivalent. Thus we arrive at
\ba\fl
\bK=\bGamma_0(-1)^{NL}(\bA_L)^N\omega^{\halfs N(N+1)L}+\nonumber\\
\fl \sum_{\ell_1=0}^L\!\vphantom{\sum}'\!\cdots\!
\sum_{\ell_N=0}^L\!\!\vphantom{\sum}'\sum_{m=0}^{L-1}
\bR_m\bq_{NL-m-\ell_1-\cdots-\ell_N-1}(-1)^{m+\ell_1+\cdots+\ell_N}
\bA_{\ell_1}\bA_{\ell_2}\cdots\bA_{\ell_N}
\omega^{\ell_2+\cdots+(N-1)\ell_N}.
\label{bKK}\ea
The summation over $\ell_1$ can be carried out using (\ref{q0})
resulting in
\ba\fl
\bK=\bGamma_0(-1)^{NL}(\bA_L)^N\omega^{\halfs N(N+1)L}+\nonumber\\
\fl \sum_{\ell_2=0}^L\cdots\sum_{\ell_N=0}^L\sum_{m=0}^{L-1}
\bR_m\delta_{NL-1,m+\ell_2+\cdots+\ell_N}(-1)^{m+\ell_2+\cdots+\ell_N}
\bA_{\ell_2}\cdots\bA_{\ell_N}\omega^{\ell_2+\cdots+(N-1)\ell_N}.
\label{Kf}\ea
There is only way for $m+\ell_2+\cdots+\ell_N=NL-1$ to hold,
namely $m=L-1$ and $\ell_2=\cdots=\ell_N=L$. From (\ref{Ljsum}) and
(\ref{Ljpm}) we can easily see that ${\hat\bC}_{L-1}=0$, and from
(\ref{A0L}) we find 
\be
{\hat\bA}_{L-1}=\prod_{j=2}^L\balpha^-_j, \quad\hbox{so that}\quad
\bR_{L-1}=\bGamma_0\bA_L,
\ee
as seen from (\ref{Rm}) and (\ref{alpha}). Consequently, (\ref{Kf})
becomes
\ba\fl
\bK=\sum_{\ell=0}^L s_\ell \bGamma_{NL-\ell N}=
\bGamma_0(-1)^{NL+NL+L}(\bA_L)^N+\bGamma_0(-1)^{NL-1+NL-L}(\bA_L)^N=0.
\ea
This proves (B4.3). In fact, for $j>0$ it is more straightforward to
prove (\ref{B4.3j}) by simply substituting (\ref{gammaj}) into the
sum and then use (\ref{q0}).
\section[5]{Explicit form of \boldmath{$\widehat\bGamma_j$}}

\subsection[5.1]{Eigenvectors of $\bH$}

It is worth noting that even though the $s_j$ given in (\ref{sj}) are
expressed in terms of operators, they are scalars as seen from (B3.13)
and (B3.14). Thus the elements of the $\bH$-matrix given in (B4.10)
and (B4.11) are also scalars. They can be rewritten as
\ba 
h_{ij} =\delta_{i+1,j} \quad \!\hbox{for $0\le i\le NL-2$,
$0\le j\le NL-1$}; \nonumber\\
h_{NL-1,jN}=-s_{L-j}/s_0, \quad h_{NL-1,m}=0\quad\!\hbox{for $m\ne Nj$}.
\label{pron}\ea
The eigenvalues of $\bH$ are given by Baxter in the text below (B4.19)
as $\lambda_i=r_k\omega^p$, ($0\le p\le N-1$), and the $r_k^N$ are the
roots of the polynomial
\be
s_0\lambda^{NL}+s_1\lambda^{NL-N}+\cdots+s_{L-1}\lambda^{N}+s_L=
s_0\prod_{j=1}^L(\lambda^{N}-r^N_j).
\label{spol}\ee
Let $\bV^{(i)}=(V^{(i)}_0,V^{(i)}_1,\cdots, V^{(i)}_{NL-1})$ denote
the eigenvector whose eigenvalue is $\lambda_i$.
Then
\be
(\bH\bV^{(i)})_j=V^{(i)}_{j+1}=\lambda_iV^{(i)}_j, \quad
\hbox{for $0\le j\le NL-2$},
\ee
so that
\be
V^{(i)}_j=\lambda^j_i, \quad\hbox{for $0\le j\le NL-1$},
\label{vi}\ee
if we choose the normalization $V^{(i)}_{0}=1$.
Consider now the last row of $\bH$ given explicitly in (B4.10)
or also (\ref{pron}). It follows that
\be\fl
(\bH\bV^{(i)})_{NL-1}=
(-s_LV^{(i)}_0-s_{L-1}V^{(i)}_N-\cdots-s_1V^{(i)}_{NL-N})/s_0
=\lambda_iV^{(i)}_{NL-1},
\ee
which is seen from (\ref{vi}) to be
\be
-s_L-s_{L-1}\lambda_i^N-\cdots-s_1\lambda_i^{NL-N}=s_0\lambda_i^{NL},
\ee
consistent with (\ref{vi}) for $j=NL-1$.
Since $\lambda^N_i=r_i^N$ are the roots of the above polynomial
(\ref{spol}), we find that the $\bV^{(i)}$ with elements given by
(\ref{vi}) are indeed the eigenvectors of $\bH$.
Obviously matrix $\bP$ diagonalizing $\bH$ is a Vandermonde matrix,
and the elements of its inverse $(P^{-1})_{ik}$ are the coefficients
of the polynomials $f_i(z)$ given by
\be
f_i(z)=\prod_{j=1,j\ne i}^{NL}\frac{z-\lambda_j}{\lambda_i-\lambda_j}
=\sum_{k=0}^{NL-1}(P^{-1})_{ik}z^k, \quad \hbox{satisfying}
\quad f_i(\lambda_j)=\delta_{ij}.
\label{Pinverse}\ee
This is essentially Prony's 1795 result \cite{Prony,APsu2}. 
\subsection[5.2]{Alternative form for $\bq_\ell$}

Let $\bQ(t)$ be a polynomial given by
\be
\bQ(t)=\sum_{\ell=0}^\infty\bq_\ell(\omega t)^{\ell}.
\label{Q}\ee
Because of (\ref{q0}), we find
\be
\bA(t)\bQ(t)=\sum_{m=0}^\infty (\omega t)^m
\sum_{\ell=0}^m(-1)^\ell \bA_\ell\bq_{m-\ell}
=\sum_{m=0}^\infty (\omega t)^m\delta_{m,0}\mathbf{1}=\mathbf{1}.
\ee
Consequently, we have $\bQ(t)=\bA(t)^{-1}$. If we rewrite $\bA(t)$ as
\be
\bA(t)=\sum_{\ell=0}^L {\bf A}_\ell(-\omega t)^\ell=
A_0\prod_{\ell=1}^L(1-\omega t\bu_\ell),
\label{bu}\ee
where the $\bu_\ell$ are commuting operators, as the set of $\bA(t)$
for varying $t$ forms a commuting family. Since the eigenvalues of
$\btau_2(t)$ are given by Baxter in (B3.19) as
\be
A_0\prod_{\ell=1}^L (1-r_\ell\omega^{n_\ell+1}t),
\quad 0\le n_\ell\le N-1,
\ee 
the eigenvalues of $\bu_\ell$ are $r_\ell \omega^{n_\ell}$. 
We find 
\be\fl
\bQ(t)=\frac 1{\bA(t)}=
\sum_{\ell=1}^L \frac{\bOmega_\ell}{1-\omega t\bu_\ell}=
\sum_{\ell=1}^L\bOmega_\ell\sum_{m=0}^\infty(\omega t\bu_\ell)^m=
\sum_{m=0}^\infty(\omega t)^m\sum_{\ell=1}^L\bOmega_\ell\bu_\ell^m,
\ee
where $\bOmega_\ell$ can be easily found by the residue theorem.
This means that $\bq_m$ has an alternative expression,
\be
\bq_m=\sum_{\ell=1}^L\bOmega_\ell\bu_\ell^m,\quad \bOmega_\ell=
\Bigg[A_0\prod_{n=1,n\ne\ell}^L(1-\bu_n/\bu_\ell)\Bigg]^{-1}.
\label{qal}\ee
It can be shown as was done in our previous work \cite{APsu2,APsu4}
that this expression for $\bq_m$ is valid for all $m>-L$, and is
identically 0 when $-L<m<0$, see particularly eqs.~(50) of \cite{APsu2}
and (71) of \cite{APsu4} and nearby text.
\subsection[5.3]{Explicit Form of $\widehat\bGamma_j$}

From the first equality in (\ref{gammaj}) and definition (B4.17),
we find
\be
\widehat\bGamma_i=\sum_{j=0}^{NL-1}P^{-1}_{ij}\bGamma_j
=\sum_{m=0}^{L-1}\bR_m(-1)^m 
\sum_{j=0}^{NL-1}P^{-1}_{ij}\omega^j\bq_{j-1-m},
\label{widehatGamma}\ee
in which the elements of $\bP^{-1}$ are given in (\ref{Pinverse}),
and $\bq_{j-1-m}=0$ for $j-1<m<L$ as was said below (\ref{qal}).
We follow the convention of Baxter to denote the $i$th eigenvalues
of $\bH$ by $\lambda_{p,k}=r_k\omega^p$, i.e.\ identifying $i=(p,k)$,
and substitute (\ref{qal}) into the above equation to obtain
\be
\widehat\bGamma_{p,k}=\sum_{m=0}^{L-1}\bR_m(-1)^m 
\sum_{j=0}^{NL-1}P^{-1}_{p,k;j}\omega^j
\sum_{\ell=1}^L\bOmega_\ell\bu_\ell^{j-1-m}.
\label{hgamma}\ee
Unlike the $\bu_\ell$, the elements of $\bP^{-1}$ given in
(\ref{Pinverse}) are scalars multiplied by the unit operator, thus
they commutes with all other operators. We denote the eigenvectors
of the Hamiltonian $\cal H$ by
$|\{n_i\}\rangle=|n_1,\cdots, n_k,\cdots, n_L\rangle$ such that,
as in  (B.4.23),
\be
\mathcal{H}|\{n_i\}\rangle=-\sum_{j=1}^{L}\omega^{n_j} r_j|\{n_i\}\rangle,
\quad\hbox {and}\quad
\bu_\ell|\{n_i\}\rangle= r_\ell\omega^{n_\ell}|\{n_i\}\rangle.
\label{Hami}\ee 
We also rewrite (\ref{Pinverse}) as
\be
f_{p,k}( r_\ell\omega^{n_\ell})=\sum_{j=0}^{NL-1}P^{-1}_{p,k;j}
(r_\ell\omega^{n_\ell})^j=\delta_{k,\ell}\delta_{p,n_k}.
\ee 
Consequently, (\ref{hgamma}) becomes
\ba\fl
\widehat\bGamma_{p,k}|\{n_i\}\rangle
&=&\sum_{m=0}^{L-1}\sum_{\ell=1}^L(-1)^m\bR_m\bOmega_\ell
\bu_\ell^{-1-m}|\{n_i\}\rangle 
\delta_{k,\ell}\delta_{n_k+1,p}\nonumber\\
&=&\delta_{n_k,p-1}\Omega_{p-1,k}\sum_{m=0}^{L-1}(-1)^m
\bR_m(\omega^{p-1}r_k)^{-1-m}|\{n_i\}\rangle, 
\label{Gright}\ea
where
\be
\fl\bOmega_k|\{n_i\}\rangle=\Omega_{p-1,k}|\{n_i\}\rangle,\qquad
\Omega_{p-1,k} =
1\Bigg/\bigg[A_0\prod_{i=1,i\ne k}^L(1-\omega^{n_i-p+1}r_i/r_k)\bigg].
\label{Omega}\ee
Similarly, we may use  the second formula in (\ref{gammaj}) to obtain
\ba
\langle\{n_i\}|\widehat\bGamma_{p,k}
=\omega\delta_{n_k,p}\Omega_{p,k}\sum_{m=0}^{L-1}(-1)^m
\langle\{n_i\}|\bR_m(r_k\omega^p)^{-1-m}.
\label{leftG}\ea
These results are in agreement with (B4.25) and surrounding text,
\ba\fl
\widehat\bGamma_{p,k}|\{n_i\}\rangle=
\widehat\bGamma_{p,k}|n_1,\cdots,{\overset {k}{p-1}},\cdots,n_L\rangle
=\Lambda_{p,k}(\{n_i\})|n_1,\cdots,{\overset {k}{p}},\cdots,n_L\rangle, 
\label{Gright2}\ea
where $\Lambda_{p,k}$ depends on $p$ and also on $n_i$ for $i\ne k$.
More precisely, $\Lambda_{p,k}$ is given, either by (\ref{Gright})
or by (\ref{leftG}), as
\ba\fl
\Lambda_{p,k}(\{n_i\})&=&(r_k\omega^{p-1})^{-1}\Omega_{p,k}
\langle n_1,\cdots,{\overset {k}{p}},\cdots, n_L|\bY(r_k\omega^{p})|n_1,
\cdots,{\overset {k}{p-1}}\cdots,n_L\rangle\label{Lambda1}\\
\fl &=&(r_k\omega^{p-1})^{-1}\Omega_{p-1,k}\langle n_1,\cdots,
{\overset {k}{p}},\cdots, n_L|\bY(r_k\omega^{p-1})|n_1,\cdots,
{\overset {k}{p-1}}\cdots,n_L\rangle.
\label{Lambda2}\ea
in which
\ba
\bY(z)\equiv\sum_{m=0}^{L-1}(-1)^m\bR_m z^{-m}.\ea
From (\ref{Gright2}), we find that the $\widehat\bGamma_{p,k}$ behaves
as cyclic raising operators. We shall now simplify the constant
$\Lambda_{p,k}(\{n_i\})$.

\subsection[5.4]{Simplification of $\Lambda_{p,k}(\{n_i\})$}

Since $\mathcal{H}=-\bA_1/A_0$, we may use (\ref{comR1}) to find
\ba\fl
\bY(z)\mathcal{H}-\mathcal{H}\bY(z)=(\omega-1)\sum_{m=0}^{L-1}(-z)^{-m}
\Bigg[\bR_0\frac{\bA_{m+1}}{A_0}-\bR_{m+1}\Bigg]\nonumber\\
\fl=(1-\omega)z\sum_{m=1}^{L}(-z)^{-m}
\Bigg[\bR_0\frac{\bA_{m}}{A_0}-\bR_{m}\Bigg]
=(1-\omega)z\bigg[\bR_0\sum_{m=0}^{L}(-z)^{-m}
\frac{\bA_{m}}{A_0}-\bY(z)\Bigg],
\ea
where we have shifted the summation index by one, then used the fact
that $\bR_L=0$ and finally extended the summation to include $m=0$,
as the zeroth term in the sum also vanishes identically. Now we can
use (\ref{bu}) and (\ref{Hami}) to rewrite the above equation as
\ba\fl
\langle\{n'_i\}|\bY(z)|\{n_i\}\rangle\bigg[z(1-\omega)-
\sum_{i=1}^L r_i(\omega^{n_i}-\omega^{n'_i})\bigg]\nonumber\\
=z(1-\omega)\langle\{n'_i\}|\bR_0|\{n_i\}\rangle
\prod_{i=1}^L(1-\omega^{n_i}r_i/z).
\ea
If we let $z=\omega^{n_k}r_k$, the right-hand side is identically zero.
Then, identically to what Baxter did, we find that for
$\langle\{n'_i\}|\bY(\omega^{n_k}r_k)|\{n_i\}\rangle$ to be non-zero,
we must have $n'_i=n_i$ for $i\ne k$ and $n'_k=n_k+1$.

Therefore, for $z=\omega^{n_k+1}r_k$, with $n'_i=n_i$ for $i\ne k$,
and $n'_k=n_k+1$, we find
\ba\fl
\langle\{n'_i\}|\bY(\omega^{n_k+1}r_k)|\{n_i\}\rangle=
\langle\{n'_i\}|\bR_0|\{n_i\}\rangle
\prod_{i=1,i\ne k}^L(1-\omega^{n_i-n_k-1}r_i/r_k).
\ea
Letting $n_k=p-1$, and comparing the above equation
with (\ref{Lambda1}) and (\ref{Omega}), we find
\ba\fl
\Lambda_{p,k}(\{n_i\})=A_0^{-1}(r_k\omega^{p-1})^{-1}
\langle n_1,\cdots,{\overset {k}{p}},\cdots, n_L|\bR_0|n_1,
\cdots,{\overset {k}{p-1}}\cdots,n_L\rangle.
\label{Lambda3}\ea
Now (\ref{comR2}) can be used to show that (\ref{Lambda2}) can be
simplified to yield the identical result.
From (\ref{gamma1}), (\ref{Rm}) and (\ref{gamma0j}), we find
\ba \omega A_0^{-1}\bR_0=\bGamma_1=
(\omega^{-1}-1)^{-1}(\mathcal{H}\bGamma_0-\bGamma_0\mathcal{H}),
\ea
so that (\ref{Lambda3}) can be even further simplified to
\ba\fl
\Lambda_{p,k}(\{n_i\})=\langle \{n'_i\}|\bGamma_0| \{n_i\}\rangle=
\langle n_1,\cdots,{\overset {k}{p}},\cdots, n_L|\bGamma_0|n_1,
\cdots,{\overset {k}{p-1}}\cdots,n_L\rangle.
\label{Lambda4}\ea

Thus to prove (B5.4), we need to prove
\ba\fl
(r_k\omega^{p-1}-r_\ell\omega^q)
\langle n_1,\cdots,{\overset {k}{p}},\cdots, {\overset {\ell}{q}},
\cdots,n_L|\bGamma_0|n_1,\cdots,{\overset {k}{p-1}},\cdots,
{\overset {\ell}{q}},\cdots,n_L\rangle
\nonumber\\
\langle n_1,\cdots,{\overset {k}{p-1}},\cdots,
{\overset {\ell}{q}},\cdots,n_L|\bGamma_0|n_1,\cdots,
{\overset {k}{p-1}},\cdots, {\overset {\ell}{q-1}},\cdots,n_L\rangle
+\nonumber\\\fl
(r_\ell\omega^{q-1}-r_k\omega^p) \langle n_1,\cdots,{\overset {k}{p}},
\cdots,{\overset {\ell}{q}},\cdots,n_L|\bGamma_0|n_1,\cdots,
{\overset {k}{p}},\cdots, {\overset {\ell}{q-1}},\cdots,n_L\rangle
\nonumber\\
\langle n_1,\cdots,{\overset {k}{p}},\cdots,
{\overset {\ell}{q-1}},\cdots,n_L|\bGamma_0|n_1,\cdots,
{\overset {k}{p-1}},\cdots, {\overset {\ell}{q-1}},\cdots,n_L\rangle=0,
\label{b5.4}\ea
which we have not yet succeeded in doing.
\section{Summary}

Let us now summarize the main steps in  our proof of the conjectures
of Baxter. As the first $\bGamma$ in \cite{BaxterPf} is
$\bGamma_0=\bZ_1^{-1}$ in (B4.1), we split in (\ref{Al}) of section 2
the coefficients $\bA_\ell$ in the expansion of $\btau_2(t)$ into
hatted operators acting on sites 2 to $L$ and operators
(\ref{Gamma0}) and (\ref{beta}) acting on site 1.  Thus the hatted
operators commute with $\bGamma_0$, $\balpha_1^\pm$ and $\bbeta_1^-$.
In subsection 4.1 we give the general formula (\ref{gammaj})
for $\bGamma_j$, which we proved by induction. It was originally
discovered calculating $\bGamma_j$ for $j=1,2,3$ using (\ref{A1})
and (\ref{gamma}).

Conjecture (B4.3) is proved in subsection 4.2. We first express the
coefficients $s_j$ in terms of the $\bA_\ell$, see (\ref{sj}). We
also rewrite (\ref{q}) as (\ref{q0}), replacing the upper limits of
the sums by $L$, as $\bA_n=0$ for $n>L$ and $\bq_\ell=0$ for $\ell<0$.
Likewise, we replace the upper limit of the summation in
(\ref{gammaj}) by $L-1$, as $\bR_m=0$ for $m>L-1$. This allows us
to interchange the summations in (\ref{bKK}) and to show using
(\ref{q0}) that (B4.3) holds.

In section 3, we proved that the coefficients of the expansion of
$t\bnu_{1}=\bmu_0$ in powers of $t$ are
equal using the commutation relations and (\ref{YB1}). The proof
of (B4.8) then follows by simple repeated application of (\ref{hmn}).

In subsection 5.1, we show that the $\bP$ of (B4.16) diagonalizing
the $\bH$ of (B4.10) is a Vandermonde matrix. Its inverse is therefore
given by (\ref{Pinverse}). In subsection 5.2, we show that the
$\bq_\ell$ defined in (\ref{q}) are coefficients of the inverse
of $\bA(t)$, and thus have the alternative form (\ref{qal}).
These equations are then used in subsection 5.3 to show that the
$\widehat\bGamma_{p,k}$ when acting on the eigenvectors of the
Hamiltonian, behave as cyclic raising operators, see (\ref{Gright2}).
The proportionality constant $\Lambda_{p,k}$ in (\ref{Gright2})
is simplified in subsection 5.4. We have not yet succeeded in
proving (B5.4), but reduced it to a simpler form (\ref{b5.4}).

Since the $\btau_2(t)$ matrices considered here are most general,
it may be interesting to see what these cyclic raising operators
are in certain special cases, and to compare with Fendley's work
\cite {Fen912, Fen1212}. In particular, a proof of (B5.4) should also
provide a proof of (111) in \cite{Fen13}. From (\ref{Gright})
and (\ref{leftG}), we see that the $\hat\bGamma_j$ are cyclic raising
operators when acting on the right, and cyclic lowering operators when
acting on the left. It should be interesting to find out what these
operators do in the full integrable chiral Potts model.
\section*{Acknowledgments}

The authors thank Professor R.J.~Baxter for many helpful
discussions and valuable correspondence. They also thank Professor
P.~Fendley for some helpful discussions about the connections between
the different approaches. Appendix B was added while the
authors were visiting the Wuhan Institute of Physics and Mathematics
(WIPM) of the Chinese Academy of Sciences. The warm hospitality and
financial support during this visit is gratefully acknowledged.
\appendix
\section{Yang--Baxter Equation\label{append}}

There are many different conventions for setting up the Yang--Baxter
equation, which are slightly different, leading to different
multiplicative factors and other changes. As an example, in our previous
papers \cite{APsu2,APsu4}, we have unfortunately used the convention of
multiplying matrices from up to down, which causes $\bX\to\bX^{-1}$ as
compared to Baxter's choice. Here we shall adopt Baxter's convention.
For this reason, it may be good to provide some details of our setup
used in the main text.

\begin{figure}[!hbtp]\begin{center}
\includegraphics[width=1.\hsize]{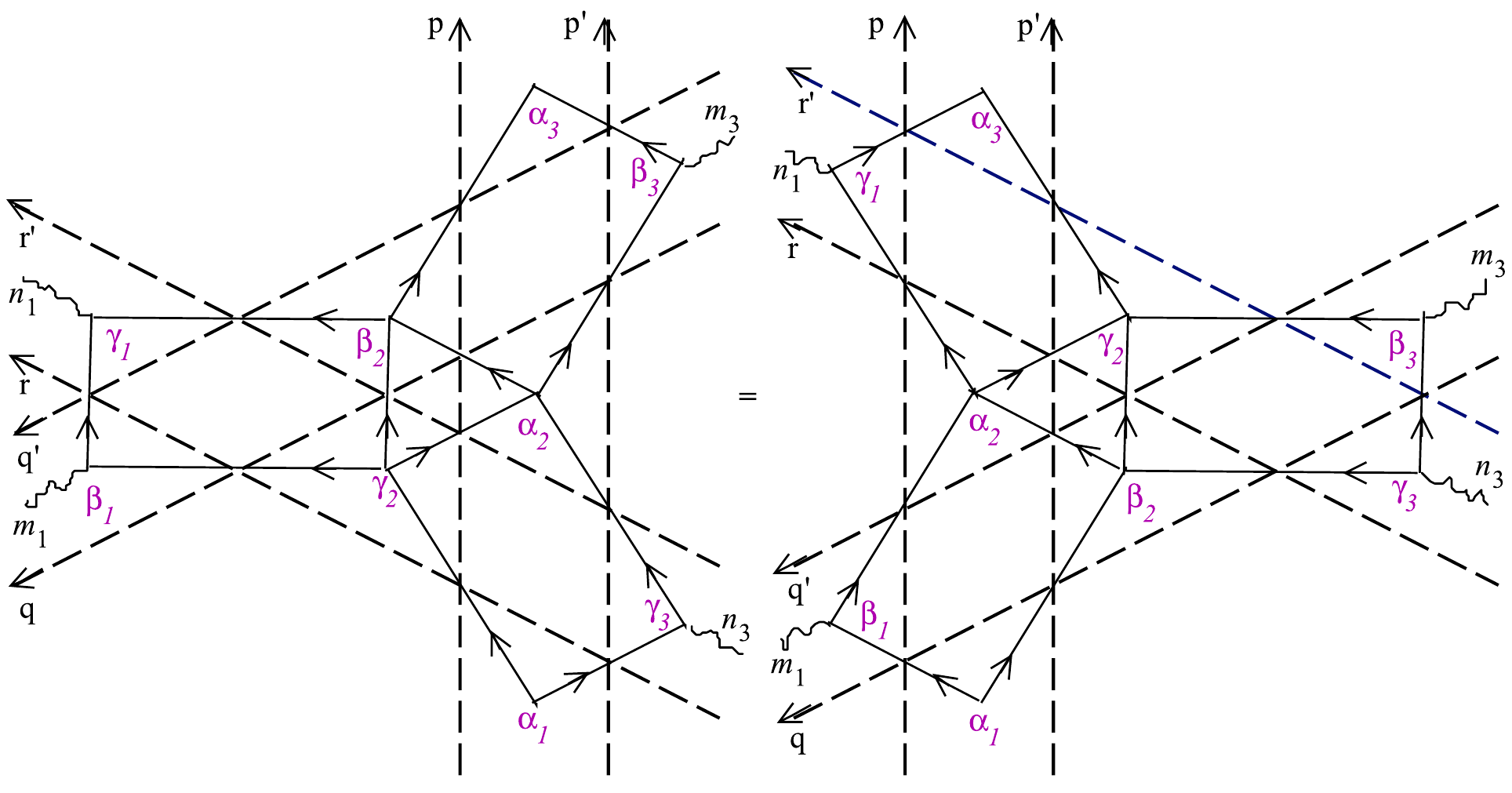}
\caption{The Yang--Baxter equation for the chiral Potts model.
The rapidity lines are represented by dashed oriented lines,
the Boltzmann weights by oriented line pieces connecting
pairs of spins.\label{fig1}}
\end{center}
\end{figure}

The products of four chiral Potts model weights \cite{BPA} satisfy
the Yang--Baxter equation 
\ba\fl
\sum_{\alpha_2,\beta_2,\gamma_2}
{\bar S}(rr'qq')_{\gamma_1,\beta_1}^{\gamma_2,\beta_2}
S(pp'rr')_{\alpha_1,\gamma_2}^{\alpha_2,\gamma_3}
S(pp'qq')_{\alpha_2,\beta_2}^{\alpha_3,\beta_3}\nonumber\\
=\sum_{\alpha_2,\beta_2,\gamma_2}
S(pp'qq')_{\alpha_1,\beta_1}^{\alpha_2,\beta_2}
S(pp'rr')_{\alpha_2,\gamma_1}^{\alpha_3,\gamma_2}
{\bar S}(rr'qq')_{\gamma_2,\beta_2}^{\gamma_3,\beta_3},
\label{YBE}\ea
as shown in figure \ref{fig1}. From the figure, we can also see
that \cite{BPA}
\be\fl
S(pp'qq')_{\alpha^{\vp},\beta}^{\alpha'\!,\beta'}=
W_{p'q}(\alpha-\beta'){\bar W}_{p'q'}(\beta'-\alpha')
{\bar W}_{pq}(\alpha-\beta)W_{pq'}(\beta-\alpha').
\label{defS}\ee
For the chiral Potts model, arrows must be drawn on the rapidity
lines and also on the line pieces representing the Boltzmann weights.
In our earlier papers, the matrix multiplications were done from up
to down in order to have all $S$ defined identically.
Here, however, doing the multiplication in the other direction,
we let
\ba\fl
{\bar S}(rr'qq')_{\gamma^{\vp},\beta}^{\gamma'\!,\beta'}
={ S}(rr'qq')_{\gamma'\!,\beta}^{\gamma^{\vp},\beta'}
=W_{r'q}(\gamma'-\beta'){\bar W}_{r'q'}(\beta'-\gamma)
{\bar W}_{rq}(\gamma'-\beta)W_{rq'}(\beta-\gamma).
\ea
Indeed, from the figure we can see that we must interchange
$\gamma$ and $\gamma'$ in order to be fully consistent with
the four arrows on the $S$ weights.

Next we multiply both sides of (\ref{YBE}) by
$\omega^{-m_1\beta_1-n_1\gamma_1+m_3\beta_3+n_3\gamma_3}$,
sum over $\beta_1$, $\beta_3$, $\gamma_1$ and $\gamma_3$,
and change one of the $\beta^{\phantom{x}}_2$ to $\beta'_2$
and one $\gamma^{\phantom{x}}_2$ to $\gamma'_2$, while
inserting
$$\delta_{\beta'_2\beta^{\phantom{x}}_2}
\delta_{\gamma'_2\gamma^{\phantom{x}}_2}
=N^{-2}\sum_{m_2,n_2}\omega^{m_2(\beta'_2-\beta^{\phantom{x}}_2)
  +n_2(\gamma'_2-\gamma^{\phantom{x}}_2)}$$
  and summing over $\beta'_2$ and $\gamma'_2$.
Then, defining the Fourier transforms
\ba
S^{(pf)}(pp'qq')_{\alpha^{\vp},m}^{\alpha'\!,m'}
&=&N^{-2}\sum_{\beta,\beta'} \omega^{-m\beta+m'\beta'}
S(pp'qq')_{\alpha^{\vp},\beta}^{\alpha'\!,\beta'},
\label{pfourier}\\
S^{(f)}(rr'qq')_{n,m}^{n'\!,m'}
&=&N^{-4}\sum_{\gamma,\gamma',\beta,\beta'}
\omega^{-n\gamma-m\beta+n'\gamma'+m'\beta'}
{\bar S}(rr'qq')_{\gamma^{\vp},\beta}^{\gamma'\!,\beta'},
\label{fourier}\ea
the Yang--Baxter equation (\ref{YBE}) becomes
\ba\fl
\sum_{\alpha_2,m_2,n_2}
{S}^{(f)}(rr'qq')_{n_1,m_1}^{n_2,m_2}
S^{(pf)}(pp'rr')_{\alpha_1,n_2}^{\alpha_2,n_3}
S^{(pf)}(pp'qq')_{\alpha_2,m_2}^{\alpha_3,m_3}
\nonumber\\
=\sum_{\alpha_2,m_2,n_2}
S^{(pf)}(pp'qq')_{\alpha_1,m_1}^{\alpha_2,m_2}
S^{(pf)}(pp'rr')_{\alpha_2,n_1}^{\alpha_3,n_2}
{S}^{(f)}(rr'qq')_{n_2,m_2}^{n_3,m_3}.
\label{YBE2}\ea
As in \cite[eq.~(2.25)]{BBP}, we define
\be
V_{pqq'}(\alpha,\alpha';m)=
N^{-1}\sum_{\beta} \omega^{m\beta}W_{pq}(\alpha-\beta)
{\bar W}_{pq'}(\beta-\alpha'),
\ee
so that (\ref{pfourier}) becomes
\be
S^{(pf)}(p'pqq')_{\alpha^{\vp},m}^{\alpha'\!,m'}=
V_{p'qq'}(\alpha,\alpha';m')V_{pq'q}(-\alpha',-\alpha;m),
\label{Spf}\ee
whereas (\ref{fourier}) can be rewritten as
\be\fl
S^{(f)}(rr'qq')_{n^\vp,m}^{n'\!,m'}=
N^{-2}\sum_{\gamma,\gamma'} \omega^{-n\gamma+n'\gamma'}
V_{r'qq'}(\gamma',\gamma;m')V_{rq'q}(-\gamma,-\gamma';m).
\label{Sf}\ee

It has been shown in \cite{BBP} that if the rapidities $q$
and $q'$ are related by
\be
(a_{q'},b_{q'},c_{q'},d_{q'})=(b_{q},\omega^2 a_{q},d_{q},c_q),
\label{relqq'}\ee
i.e.~\cite[eq.~(2.28)]{BBP} with $k=0$ and $\ell=2$,
then $V_{p'qq'}(\alpha,\alpha';m')$ is block-triangular.
More precisely, when $0\le \alpha-\alpha'\le 1$, 
\be
V_{p'qq'}(\alpha,\alpha';m')=0,\quad \hbox{for $2\le m' \le N-1$},
\label{Vpqq'0}\ee
while, for $0\le m'\le 1$,
\ba
V_{p'qq'}(\alpha,\alpha';m')={\Omega}_{p'q}\omega^{m'\alpha'}
(b_q/d_q)^{\alpha-\alpha'}(c_q/b_q)^{m'}F_{p'q}(\alpha-\alpha',m').
\label{Vpqq'}\ea
This is precisely  \cite[eq.~(3.39a)]{BBP} after using
\cite[eq.~(3.21)]{BBP} with $k=0$ and $\ell=2$ and
$y_q=b_q/c_q$ \cite[eq.~(2.6)]{BBP}.
Under the same condition (\ref{relqq'}),
$V_{pq'q}(-\alpha'\!,-\alpha;m)$ is also found to be block-triangular,
such that, for $0\le m\le 1$,
\be
V_{pq'q}(-\alpha'\!,-\alpha;m)=0,\quad
\hbox{for $2\le \alpha-\alpha'\le N-1$},
\label{Vpq'q0}\ee
while it is non-vanishing for $0\le \alpha-\alpha'\le 1$ and
given by
\ba\fl
V_{pq'q}(-\alpha'\!,-\alpha;m)=
{\bar\Omega}_{pq}\omega^{-m\alpha}(d_q/b_q)^{\alpha-\alpha'}
(b_q/c_q)^{m}(-\omega t_q)^{\alpha-\alpha'-m}F_{pq}(\alpha-\alpha',m).
\label{Vpq'q}\ea
This follows from \cite[eq.~(3.39b)]{BBP} absorbing the factor
$\bar h^{(j)}_{pq}$ into $\bar\Omega_{pq}$, while evaluating
$\eta_{q,2,\alpha-\alpha'}/\eta_{q,2,m'}$ 
using \cite[eq.~(3.48)]{BBP}.

Consequently, we find the diagonal block of (\ref{Spf})
for $0\le m',m\le 1$ to be
\be
S^{(pf)}(pp'qq')_{\alpha^{\vp},m}^{\alpha'\!,m'}
={\bar\Omega}_{pq}{\Omega}_{p'q}(b_q/c_q)^{m-m'}
\mathcal{L}(pp'q)_{\alpha^{\vp},m}^{\alpha'\!,m'},
\label{Lpp'q}\ee
with $\mathcal{L}$ given in (\ref{Lsj}), identifying
$\mathcal{L}(pp'q)_{\alpha^{\vp},m}^{\alpha'\!,m'}=
\mathcal{L}_j(m,m';\alpha,\alpha')$ there.
In ${\bar\Omega}_{pq}{\Omega}_{p'q}$ we have collected irrelevant
factors that cancel out of the Yang--Baxter equation.
Likewise if the two rapidities $r$ and $r'$ are also related by 
\be
(a_{r'},b_{r'},c_{r'},d_{r'})=(b_{r},\omega^2 a_{r},d_{r},c_r), 
\label{relrr'}\ee
we have
\be
S^{(pf)}(pp'rr')_{\alpha^{\vp},n}^{\alpha'\!,n'}
={\bar\Omega}_{pr}{\Omega}_{p'r}(b_r/c_r)^{n-n'}
\mathcal{L}(pp'r)_{\alpha^{\vp},n}^{\alpha'\!,n'}.
\label{Lpp'r}\ee

Consider now the Fourier transform (\ref{Sf}). If $q'$ and $q$
are related by (\ref{relqq'}), we find from (\ref{Vpq'q0}) that
for $0\le m\le 1$ that $V_{rq'q}(-\gamma,-\gamma',m)$ is
non-vanishing only when $\gamma'-\gamma=0,1$. Thus, if we change
the sum over $\gamma'$ to one over $\ell=\gamma'-\gamma$ and then
sum over $\gamma$, we obtain
\ba\fl
S^{(f)}(rr'qq')_{n^\vp,m}^{n',m'}=
{\bar\Omega}_{rq}{\Omega}_{r'q}(b_q/c_q)^{m-m'}N^{-2}
\sum_{\gamma} \omega^{(-n+n'+m'-m)\gamma}
\mathcal{R}(rr'q)_{n^{\vp},m}^{n'\!,m'}\nonumber\\
={\bar\Omega}_{rq}{\Omega}_{r'q}(b_q/c_q)^{m-m'}N^{-1}
\delta_{n'+m',n+m}
\mathcal{R}(rr'q)_{n^{\vp},m}^{n'\!,m'},
\label{SR}\ea
where
\be\mathcal{R}(rr'q)_{n^{\vp},m}^{n'\!,m'}
=\sum_{\ell=0}^1 \omega^{(n'-m)\ell}
F_{r'q}(\ell,m')F_{rq}(\ell,m)(-\omega t_q)^{\ell-m}.
\ee

It is straightforward to show that when $r'$ and $r$ are also
related by (\ref{relrr'}), $\mathcal{R}(rr'q)_{n^{\vp},m}^{n'\!,m'}=0$
for $0\le n\le 1$ and $2\le n'\le N-1$, while for $0\le n,n'\le 1$
it is given by
\ba
\delta_{n'+m',n+m}\mathcal{R}(rr'q)_{n^{\vp},m}^{n'\!,m'}
=(b_r/c_r)^{m'-m}\mathcal{R}(rq)_{n^{\vp},m}^{n'\!,m'},\\
\mathcal{R}(rq)_{n^{\vp},m}^{n'\!,m'}=\delta_{n'+m',n+m}
\left[\Bigg(\frac{-t_q}{\omega t_r}\Bigg)^{m'}-(-1)^{m'}
\omega^{n-1}\Bigg(\frac{t_q}{t_r}\Bigg)^{1-m}\right].
\ea
Here we used \cite[eq.~(3.48)]{BBP} identifying
$F_{pq}(\ell,m)=F_{pq}(2,\ell,m)$, which differs from
(B2.2) used to derive (\ref{Lj}) by a normalization factor $b_p$.
In particular, we have
\ba
\mathcal{R}(rq)_{0,0}^{0,0}=
\mathcal{R}(rq)_{1,1}^{1,1}=1-t_q/(\omega t_r),\nonumber\\
\mathcal{R}(rq)_{1,0}^{1,0}=
\omega\mathcal{R}(rq)_{0,1}^{0,1}=1-t_q/ t_r,\nonumber\\
\mathcal{R}(rq)_{0,1}^{1,0}=
(t_r/t_q)\mathcal{R}(rq)_{1,0}^{0,1}=1-\omega^{-1}.
\ea
This shows that, when both relations in (\ref{relqq'})
and (\ref{relrr'}) hold, the Fourier transform of the product
of four Boltzmann weights (\ref{Sf}) reduces to the weights of a
six-vertex model given as
\ba\fl
S^{(f)}(rr'qq')_{n^\vp,m}^{n'\!,m'}=
{\bar\Omega}_{rq}{\Omega}_{r'q}(b_q/c_q)^{m-m'}
(b_r/c_r)^{n-n'}N^{-1}\mathcal{R}(rq)_{n^{\vp},m}^{n'\!,m'},
\label{SR6v}\ea
for $0\le m,n,m',n'\le 1$. Substituting (\ref{Lpp'q}), (\ref{Lpp'r})
and (\ref{SR6v}) into the Yang--Baxter equation (\ref{YBE2}),
we find that many factors cancel out leaving us with
\ba\fl
\sum_{\alpha_2,m_2,n_2}
\mathcal{R}(rq)_{n_1,m_1}^{n_2,m_2}
\mathcal{L}(pp'r)_{\alpha_1,n_2}^{\alpha_2,n_3}
\mathcal{L}(pp'q)_{\alpha_2,m_2}^{\alpha_3,m_3}
\nonumber\\
=\sum_{\alpha_2,m_2,n_2}
\mathcal{L}(pp'q)_{\alpha_1,m_1}^{\alpha_2,m_2}
\mathcal{L}(pp'r)_{\alpha_2,n_1}^{\alpha_3,n_2}
\mathcal{R}(rq)_{n_2,m_2}^{n_3,m_3}.
\label{YBE3}\ea

It is easily verified that this relation holds without
any condition on the two sets of $\{a,b,c,d\}$ parameters
making up the rapidities  $p$ and $p'$, unlike the
Yang--Baxter equation for the chiral Potts model,
for which the parameters have to satisfy
\cite[eq.~9]{BPA} defining the chiral Potts curve.
This observation has been made first by Baxter
\cite{BaxterFun} in somewhat different notations.

Finally, it is obvious, that Yang--Baxter equation (\ref{YBE3})
also holds for so-called monodromy operators (\ref{ABCD}), replacing
each $\mathcal{L}$ by a product of $\mathcal{L}$-matrices sharing a
horizontal rapidity line \cite{STF}.
In particular, letting $n_1=0$, $m_1=1$, $n_3=m_3=0$ in
(\ref{YBE3}), we obtain (\ref{YBEAC}). If we choose
$n_1=m_1=1$, $n_3=m_3=0$, we find
$\bC(x)\bC(y)=\bC(y)\bC(x)$, while using $n_1=m_1=n_3=m_3=0$, we find
$\bA(x)\bA(y)=\bA(y)\bA(x)$. Applying this to
(\ref{hatAC}) we find the commutation relations
\be
{\hat\bC}_m{\hat\bC}_n={\hat\bC}_n{\hat\bC}_m,\quad
{\hat\bA}_m{\hat\bA}_n={\hat\bA}_n{\hat\bA}_m.
\label{comCA}\ee


\section{Comparison with Fendley's paper\label{appendb}}


\subsection{Opening remarks}
Before starting the comparison with \cite{Fen13}, we must remark that
we have to follow Baxter's notations of \cite{BaxterPf}, which used
$\bGamma_0=\bZ_1^{-1}$, rather than $\bGamma_0=\bZ_1$ as used in
\cite{Fen13,Baxterham,Bax89}. This results in a spatial reflection
of the way operators are multiplied. Therefore, we multiply operators
in numerical order of site number, rather than Fendley's (and Baxter's
earlier) anti-numerical order. Furthermore, in this appendix, equations
in \cite{Fen13} will be denoted by prefacing F to their equation
numbers.


\subsection{Comparing transfer matrices}
Following (B3.25), we set $a_j\equiv0$ and $b_j\equiv1$ in (\ref{Lj}).
As we now have
\ba
&&\balpha_j^+\equiv{\bf1},\qquad\bbeta_j^+=c_{2j-2}\bZ_j^{-1},
\qquad\bgamma_j^+\equiv0,\nonumber\\
&&\balpha_j^-=d_{2j-2}d_{2j-1}\bX_j,\quad
\bbeta_j^-=c_{2j-1}\bZ_j,\quad\bgamma_j^-=c_{2j-2}c_{2j-1}{\bf1},
\ea
for the quantities defined in (\ref{Ljpm}),
(\ref{Ljsum}) simplifies to
\be
\mathcal{L}_j=\left[\begin{array}{cc}
{\bf1}&0\\
\bbeta_j^+&0\end{array}\right]
-\omega t
\left[\begin{array}{cc}
\balpha_j^-&\bbeta_j^-\\
0&\bgamma^-_j\end{array}\right],
\ee
with the special relationships
\ba
&&\balpha_j^-={\bf h}_{2j-1}\equiv d_{2j-2}d_{2j-1}\bX_j,
\nonumber\\
&&\bbeta_j^-\bbeta_{j+1}^+={\bf h}_{2j}\equiv
c_{2j-1}c_{2j}\bZ_j\bZ_{j+1}^{-1},
\qquad
\bbeta_j^-\bgamma_{j+1}^-={\bf h}_{2j}\bbeta_{j+1}^-.
\ea
Noting that $\btau_2(t)$ is defined in (\ref{ABCD}) and
(\ref{A}) as the 1-1 matrix element of $\prod_\ell\mathcal{L}_\ell$,
it is then easily seen that the ${\bf A}_\ell$ also defined in
(\ref{A}) are expressed as sums of products of factors
$\balpha_j^+={\bf 1}$, $\balpha_j^-={\bf h}_{2j-1}$ and
\be
\bbeta_j^-\Bigg(\prod_{i=j+1}^{k}\bgamma_i^-\Bigg)\bbeta_{k+1}^+
=\prod_{i=j}^{k}{\bf h}_{2i}.
\ee
As we need precisely one of $\balpha_j^\pm$, $\beta_j^\pm$, or
$\bgamma_j^-$ for each site $j$, one can easily verify that we
get the exclusion rule of (F41) that the subscripts
of the ${\bf h}_j$ must be at least two apart.
More precisely, we find, in agreement with \cite{Fen13}, that
\be
{\bf A}_m=\sum_{i_1=1}^{2L-2m+1}\,\sum_{i_2=i_1+2}^{2L-2m+3}\cdots
\sum_{i_m=i_{m-1}+2}^{2L-1}\,\prod_{j=1}^m{\bf h}_{i_j},
\ee
with the special cases
\be
{\bf A}_0={\bf1},\qquad {\bf A}_L=\prod_{i=1}^{L}{\bf h}_{2i-1}.
\ee
Therefore, for this special case, we have the following relation
\be
T(-\omega t)=\btau_2(t),
\label{Tbtau}\ee
with $T(t)$ defined in (F50).


\subsection{Comparing Hamiltonians}
Next, as $\bgamma_j^+=0$, only the term with $m=j+1<L$ survives
within (\ref{H}), so that now
\be
\mathcal{H}=-\sum_{i=1}^{2L-1}{\bf h}_i,
\ee
in agreement with (F34) and (F38) (up to a trivial
minus sign) and with (B1.5) for this special case.


\subsection{Comparing the eigenvalues of the Hamiltonian}
The eigenvalues of the Hamiltonian are given in (\ref{Hami}) in terms
of $r_k\omega^{n_k}$. This has to be identified with
$\epsilon_k\omega^{n_k}$ from the action of $\mathcal{H}$ on the
cyclic raising (shift) operator in (F95); one may also look at (F102)
for $m=1$. From (F48) and (\ref{Tbtau}), we find 
\be
\sum_{j=0}^{N-1}t\frac {\rm d} {{\rm d}t}\ln[\tau_2(\omega^{j}t)]
=\sum_{s=1}H^{(sN)}t^{sN}.
\ee
Comparing (F63) and the equation above (F65) with (\ref{prodsum}) and
(\ref{spol}) in the present paper, we can see that
$u_k=\epsilon_k^N=r_k^N$, so that the eigenvalues indeed agree.

\subsection{Relation between the inverses of Vandermonde matrices}
In (\ref{Pinverse}), we have expressed the elements $(P^{-1})_{ij}$
of the inverse of the Vandermonde matrix as coefficients
of the polynomials $f_i(z)$. Identifying $i$ in (\ref{Pinverse})
with $(p,k)$ and $j$ with $(q,\ell)$ as is done in the above subsection
(5.3), this can also be rewritten as
\ba
f_{p,k}(z)&&=\prod_{\ell=1,\ell\ne k}^{L}\prod_{q=1}^N
\frac{z-r_\ell\omega^q}{r_k\omega^p-r_\ell\omega^q}
\prod_{q=1}^{N-1}\frac{z-r_k\omega^{p+q}}{r_k\omega^p(1-\omega^q)}
\nonumber\\
&&=
\frac 1N\prod_{\ell=1,\ell\ne k}^{L}\frac{z^N-r_\ell^N}{r^N_k-r^N_\ell}
\left[\frac{(z/r_k)^N-1}{z/(r_k\omega^p)-1}\right].
\label{PX}\ea
Obviously, we may express the product on the second line of
(\ref{PX}) as the inverse of the Vandermonde matrix ${\cal X}$
defined above (F100), and we may expand the part within the square
brackets as a geometric series. Then, equating coefficients, we find
\be
P_{i,\ell N+q}^{-1}=
\frac 1N({\cal X}^{-1})_{k,\ell}(r_k\omega^p)^{-q},\quad i\equiv(p,k).
\ee
Consequently, we may rewrite (\ref{widehatGamma}) as
\be\fl
\widehat\bGamma_{p,k}\equiv\sum_{j=0}^{NL-1}P^{-1}_{ij}\bGamma_j
=\sum_{s=0}^{N-1}(r_k\omega^p)^{-s}\Phi_k^{(s)}(r_k^N),
\quad \Phi_k^{(s)}(r_k^N)\equiv
\frac 1N\sum_{\ell=0}^{L-1}({\cal X}^{-1})_{k,\ell}\bGamma_{\ell N+s}.
\ee
It is easily verified that these functions $\Phi_k^{(s)}(r_k^N)$
satisfy the unnumbered relation below (F94) and the first unnumbered
equation in section 5.3 of Fendley's paper and
therefore
\be
\widehat\bGamma_{p,k}=\Psi_{\omega^p,k}
\ee
with $\Psi_{\omega^p,k}$  defined in (F95).



\section*{References}
\begin{thebibliography}{000}

\bibitem{Green}{%
Green H S 1953
\textrm{A generalized method of field quantization}
\textit{Phys. Rev.} \textbf{90} 270--3}
%
\bibitem{Bethe}{%
Bethe H 1931
\textrm{Zur Theorie der Metalle.
I. Eigenwerte und Eigenfunktionen der linearen Atomkette}
\textit{Z. Phys.} \textbf{71} 205--26}
%
\bibitem{Morris1}{%
Morris A O 1967
\textrm{On a generalized Clifford algebra}
\textit{Quart. J. Math.} \textbf{18} 7--12}
%
\bibitem{Morris2}{%
Morris A O 1968
\textrm{On a generalized Clifford algebra (II)}
\textit{Quart. J. Math.} \textbf{19} 289--99}
%
\bibitem{Morris3}{%
Morris A O 1970
\textrm{Generalized Clifford algebras and $L$-matrix hierarchy}
\textit{J. Math. Anal. Appl.} \textbf{31} 136--9}
%
\bibitem{Yamazaki}{%
Yamazaki K 1964
\textrm{On projective representations and ring extensions
of finite groups}
\textit{J. Fac. Sci. Univ. Tokyo} Sect 1 \textbf{10} 147--95
\hfill\break (Online at
http://repository.dl.itc.u-tokyo.ac.jp/dspace/handle/2261/6042 $\!$)}
%
\bibitem{PG}{%
Popovici I and Gh\'eorghe C 1966
\textrm{Alg\`ebres de Clifford g\'en\'eralis\'ees}
\textit{C. R. Acad. Sc. Paris, S\'erie A--B} \textbf{262} A682--5
\hfill\break (Online at
http://gallica.bnf.fr/ark:/12148/bpt6k6413222n/f144.image.langEN $\!$)}
%
\bibitem{Weyl}{%
Weyl H 1927
\textrm{Quantenmechanik und Gruppentheorie}
\textit{Z. Phys.} \textbf{46} 7--46, see p.~32}
%
\bibitem{Syl}{%
Sylvester J J 1883
\textrm{On quaternions, nonions, sedenions, etc.}
\textit{Johns Hopkins University Circulars}
{\textbf 3} No.~27, 7-9 (Online at
https://jscholarship.library.jhu.edu/handle/1774.2/32855 $\!$)}
%
\bibitem{FK}{%
Fradkin E and Kadanoff L P 1980
\textrm{Disorder variables and para-fermions in two-dimensional
statistical mechanics}
\textit{Nucl. Phys. B} \textbf{170}[FS1] 1--15}
%
\bibitem{KC}{%
Kadanoff L P and Ceva H 1971
\textrm{Determination of an operator algebra for
the two-dimensional Ising model}
\textit{Phys. Rev. B} \textbf{3} 3918--39}
%
\bibitem{Fen912}{%
Fendley P 2012
\textrm{Parafermionic edge zero modes in
$\mathbb{Z}_n$-invariant spin chains}
\textit{J. Stat. Mech.} 2012 P11020 (25 pp)
(arXiv:1209.0472)}
%
\bibitem{Fen1212}{%
Fendley P 2012
\textrm{Parafermions and the integrable Potts chains}
\textit{unpublished draft}}
%
\bibitem{Fen13}{%
Fendley P 2014
\textrm{Free parafermions}
\textit{J. Phys. A: Math. Theor.} \textbf{47} 075001 (42pp)
(arXiv:1310.6049)}
%
\bibitem{Baxterham}{%
Baxter R J 1989
\textrm{A simple solvable $Z_N$ Hamiltonian}
\textit{Phys. Lett. A} \textbf{140} 155--7}
%
\bibitem{Bax89}{%
Baxter R J 1989
\textrm{Superintegrable chiral Potts model: Thermodynamic properties,
an ``inverse'' model, and a simple associated Hamiltonian}
\textit{J. Stat. Phys.} \textbf{57} 1--39}
%
\bibitem{BPA}{%
Baxter R J, Perk J H H and Au-Yang H 1988
\textrm{New solutions of the star-triangle relations for the chiral
Potts model}
\textit{Phys. Lett.} A \textbf{128} 138--42}
%
\bibitem{Kri81}{%
Krichever I M 1981
\textrm{Baxter's equations and algebraic geometry}
\textit{Funkts. Anal. Prilozhen.} \textbf{15} 22--35
[\textit{Funct. Anal. Appl.} \textbf{15} 92--103]}
%
\bibitem{Kri82}{%
Krichever I M 1982
\textrm{Algebraic geometry methods in the theory of
Baxter--Yang equations}
\textit{Soviet Scientific Reviews. Section C} vol 3
(Harwood Academic Pub, Switzerland) pp 53--81}
%
\bibitem{Kor86}{%
Korepanov I G 1986
\textrm{The method of vacuum vectors in the theory of
Yang--Baxter equation}
\textit{Applied Problems in Calculus}
(Publishing House of Chelyabinsk Polytechnical Institute,
Chelyabinsk, Russia) pp 39--48
(arXiv:nlin/0010024) [This paper does the case $N=3$.
The addendum in the arXiv translation cites three related
preprints in Russian deposited in the VINITI arXiv in 1986
and 1987 also doing the case of general $N$. These can be
more easily accessed at http://yadi.sk/d/TYQ2iwL4QgJWa.]}
%
\bibitem{BS}{%
Bazhanov V V and Stroganov Yu G 1990
\textrm{Chiral Potts model as a descendant of the six-vertex model}
\textit{J. Stat. Phys.} \textbf{59} 799--817}
%
\bibitem{BBP}{%
Baxter R J, Bazhanov V V and Perk J H H 1990
\textrm{Functional relations for transfer matrices of the
chiral Potts model}
\textit{Int. J. Mod. Phys.} B \textbf{4} 803--70}
%
\bibitem{AMPTY}{%
Au-Yang H, McCoy B M, Perk J H H, Tang S and Yan M-L 1987
\textrm{Commuting transfer matrices in the chiral Potts models:
Solutions of the star-triangle equations with genus $> 1$}
\textit{Phys. Lett.} A \textbf{123} 219--23}
%
\bibitem{BaxterPf}{%
Baxter R J 2013
\textrm{The $\tau_2$ model and parafermions}
\textit{arXiv:1310.7074}}
%
\bibitem{BaxterFun}{%
Baxter R J 2004
\textrm{Transfer matrix functional relations for the
generalized $\tau_2(t_q)$ model}
\textit{J. Stat. Phys.} \textbf{117} 1--25
(arXiv:cond-mat/0409493)}
%
\bibitem{Prony}{%
Prony (G C F M R de) 1795
\textrm{Consid\'erations sur les principes de la m\'ethode
inverse des diff\'erences}
\textit{J. de l'\'Ec. Polyt.} \textbf{1} (3) 209--73, see pp 264--5
\hfill\break (Online at
http://gallica.bnf.fr/ark:/12148/bpt6k4336621/f23.image $\!$)}
%
\bibitem{APsu2}{%
Au-Yang H and Perk J H H 2009
\textrm{Eigenvectors in the superintegrable model II:
ground-state sector}
\textit{J. Phys. A: Math. Theor.} \textbf{42} 375208 (16pp)
(arXiv:0803.3029)}
%
\bibitem{APsu4}{%
Au-Yang H and Perk J H H 2011
\textrm{Quantum loop subalgebra and eigenvectors of the
superintegrable chiral Potts transfer matrices}
\textit{J. Phys. A: Math. Theor.} \textbf{44} 025205 (26pp)
(arXiv:0907.0362)}
%
\bibitem{STF}{%
Sklyanin E K, Takhtadzhyan L A and Faddeev L D 1979
\textrm{Quantum inverse problem method.~I}
\textit{Teor. Mat. Fiz.} \textbf{40} 194--220
[\textit{Theor. Math. Phys.} \textbf{40} 688--706]}

\end{thebibliography}
\end{document}